\documentclass[12pt, draftclsnofoot, onecolumn]{IEEEtran}

\usepackage{algorithm}
\usepackage{algpseudocode}
\usepackage{amsbsy}
\usepackage{amsmath}
\usepackage{amssymb}
\usepackage{bm}
\usepackage{cite}
\usepackage{graphicx}
\usepackage{mathbbol}
\usepackage{theorem}
\usepackage{url}

\newcommand{\insertfig}[2]{
\begin{figure}\centering
\includegraphics[width=.9\columnwidth]{#1.pdf}
\caption{#2}\label{#1.fig}\end{figure}}
\makeatletter
\renewenvironment{subequations}{%
  \refstepcounter{equation}%
  \protected@edef\theparentequation{\theequation}%
  \setcounter{parentequation}{\value{equation}}%
  \setcounter{equation}{0}%
  \def\theequation{\theparentequation-\alph{equation}}%
  \ignorespaces
}{%
  \setcounter{equation}{\value{parentequation}}%
  \ignorespacesafterend
}
\makeatother
\newtheorem{corollary}{Corollary}

\newtheorem{theorem}{Theorem}
\theorembodyfont{\rmfamily}

\newtheorem{remark}{Remark}
\def\ben{\begin{enumerate}}
\def\beq{\begin{equation}}
\def\beqa{\begin{eqnarray}}
\def\bit{\begin{itemize}}
\def\een{\end{enumerate}}
\def\eeq{\end{equation}}
\def\eeqa{\end{eqnarray}}
\def\eit{\end{itemize}}

\def\non{\nonumber\\}

\DeclareMathAlphabet{\mathsfbf}{OT1}{cmss}{sbc}{n}

\def\Am{\bm{A}}

\def\av{\bm{a}}

\def\Bm{\bm{B}}

\def\CN{\mathcal{CN}}

\def\Dmc{\check{\Dm}}
\def\Dm{\bm{D}}

\def\diag{{\mathsf{diag}}}

\def\e{\mathsf{e}}

\def\Fm{\bm{F}}

\def\H{\mathsfbf{H}}

\def\Hm{{\bm{H}}}

\def\hv{{\bm{h}}}

\def\Id{\bm{I}}
\def\j{\,\mathrm{j}\,}

\def\Km{\bm{K}}

\def\L{\mathcal{L}}

\def\Lam{\bm{\Lambda}}

\def\LamF{\Lam_\mathsf{F}}
\def\LamH{\Lam_\mathsf{H}}
\def\LamM{\Lam_\mathsf{M}}
\def\LamQ{\Lam_\mathsf{Q}}
\def\LamR{\Lam_\mathsf{R}}

\def\Mmm{\bm{M}}
\def\Mm{\Mmm}

\def\nR{{\mathsf{n_R}}}
\def\nT{{\mathsf{n_T}}}

\def\Pm{\bm{P}}

\def\Qm{\bm{Q}}

\def\Qmt{\tilde{\Qm}}
\def\qv{\bm{q}}

\def\rank{\mathsf{rank}}
\def\Rm{\bm{R}}

\def\T{\mathsfbf{T}}

\def\tr{\mathop{\mathsf{tr}}}

\def\U{\mathcal{U}}
\def\Um{\bm{U}}

\def\UmH{\Um_\mathsf{H}}
\def\uv{\bm{u}}

\def\Vm{\bm{V}}

\def\VmH{\Vm_\mathsf{H}}
\def\vv{\bm{v}}

\def\wv{\bm{w}}

\def\Xm{\bm{X}}

\def\xv{\bm{x}}

\def\yv{\bm{y}}

\def\0{\bm{0}}
\def\1{\bm{1}}

\def\zv{\bm{z}}

\begin{document}

\title{MIMO Capacity with Average Total and Per-Antenna Power Constraints}
\author{Giorgio Taricco}

\maketitle
\begin{abstract}
MIMO capacity with a joint total and per-antenna average power constraint is considered in this work.
The problem arises when, besides having a limited available power at the transmitter, also the individual antennas cannot radiate power beyond the limits of their corresponding RF chains.
Closed-form results are illustrated in specific cases and, in particular, in the unit-rank channel matrix case.
Lower-complexity optimization problems are derived for the other cases.
Numerical complexity is derived in terms of the number of equivalent real variables for the optimization problem and a general formula is provided depending on the channel matrix rank.
Numerical results are included to validate and illustrate the application of the proposed optimization algorithms and also to evaluate the time complexity of its implementation.
\end{abstract}


\section{Introduction}

The capacity of multipe-input multiple-output (MIMO) channels is commonly evaluated under a total average power constraint~\cite{cover,tse,bps98} in order to account for the limited availability of transmission power to be sent to the transmitting antennas.
Under the assumption of known channel matrix at both the transmitter and receiver endpoints, i.e., with perfect Channel State Information at the Transmitter (CSIT) and at the Receiver (CSIR), the capacity is obtained by applying the water-filling algorithm~\cite{cover} to the equivalent eigen-channels describing the MIMO channel itself.
The resulting solution is very simple and elegant and its derivation is quite straightforward after the application of a singular value decomposition (SVD) to the channel matrix itself.
The water-filling algorithm implementation is very simple and efficient and it solves completely the optimization problem behind the evaluation of the channel capacity.
However, some implementations of MIMO transmitters rely on separate RF chains feeding the different transmit antennas so that the average transmit power from every antenna is subject to a specific constraint~\cite{loyka17,vu11}.
Also in he case of distributed MIMO systems, where transmitting antennas are located at different places, power limitations affect different antennas separately and there is no overall power constraint~\cite{vu11}.
These scenarios are opposite, for example, to the case of an implementation based on a common amplifier followed by a passive beamforming network.
To summarize, power limitation on a MIMO transmitter can be of two different types:
$i)$ total average power (TP) constrained or $ii)$ per-antenna average power (PAP) constrained.
We can also figure out a hybrid situation where both constraints have to be enforced (TP and PAP): the individual antenna power amplifiers may trade-off some of their available power or may drain their power supply from a common source~\cite{loyka17,kl12}.
This motivates the study of the MIMO channel capacity under TP and PAP constraints.

The existing literature offers many works on the MIMO channel capacity with TP and/or PAP constraints.
One of the earliest results is due to Vu, who developed an iterative algorithm for the PAP-only MIMO case in~\cite{vu11,vupatent}.
Her approach is based on the solution of the KKT equations relevant to the PAP constrained optimization problem addressing the maximization of the MIMO channel's mutual information.
The approach is ingenious and relies on the particular structure of the optimizing covariance matrix, which is split into different signature components leading to an efficient iterative algorithm.
Nevertheless, the TP constraint is not considered and there is limited discussion of complexity and convergence issues.
More recently, in~\cite{tuni14}, the PAP-only constrained MIMO capacity has also been considered and a closed-form solution for the optimal input covariance matrix has been derived when the channel matrix and the optimal input covariance matrix have full rank.
This result was anticipated in~\cite[eq.~(23)]{vu11} in an equivalent form without giving the explicit conditions required.
Finally, Loyka~\cite{loyka17} and~\cite{cao16} considered the joint TP and PAP constrained capacity problem in the multiple-input single-output (MISO) case and obtained a closed-form solutions for the capacity-achieving input covariance matrix in this case.

In this work we address the evaluation of the capacity of a MIMO channel with perfect CSIT/CSIR under a joint TP and PAP constraint.
The paper extends in a nontrivial way the results of~\cite{vu11} and provides an algorithmic solution for this problem, which is still open for MIMO channels under a joint TP and PAP constraint (see~\cite{loyka17}).
The contributions of the paper can be summarized as follows.

We emphasize the different type of optimization problems arising when the channel matrix has full rank (equal to the number of transmit antennas) or not.
We refer to the former as the \emph{full-rank} case and to the latter as the \emph{singular} case.
In both cases we show that the capacity-achieving covariance matrix has a specific form depending on a number of real positive parameters not greater than the number of transmit antennas and on the Gramian of the channel matrix.
Moreover, we show that in the full-rank case the TP constraint is active at the optimum (i.e., the constraint is met with equality) while it may be inactive in the singular case.

The solution of the KKT equations leading to the main results discussed above is different from~\cite{vu11} for several reasons:
$i)$ the Lagrangian function is extended to encompass the TP constraint;
$ii)$ we provide a robust justification of the positivity of the diagonal Lagrange multiplier matrix $\Dm$, corresponding to the set of real positive parameters determining the optimum covariance matrix.%
\footnote{
The argument used in~\cite{vu11} to show that the Lagrange multipliers $d_i$ are real and positive is incorrect.
The paper claims that the positivity of the $d_i$'s derives from the complementary slackness condition and the fact that the optimum solution corresponds to active constraints (inequalities met as equalities).
However, if $d_i\ge0$, $d_i((\Qm)_{ii}-P_i)=0$, and $(\Qm)_{ii}=P_i$, we cannot be sure that $d_i>0$ because $d_i=0$ is also compatible with all the equations.
This property is crucial to the existence of $\Dmc\triangleq\Dm^{-1}$.
}
$iii)$ we simplify the derivation of the optimality conditions and obtain a specific relationship between the matrix $\Fm$ to the matrix $\Rm$ (matrix notation is consistent with~\cite{vu11}).

The derivation of a specific structure of the capacity-achieving covariance matrix is a key contribution of this work.
This reduces dramatically the complexity of the optimization problem.
In fact, we are considering the minimization of a strictly convex function (the negative mutual information) over a convex set (intersection of the TP, PAP, and positive semidefiniteness constraints on the input covariance matrix), which corresponds to a \emph{convex optimization problem}.
As such, its solution is global and unique~\cite{boyd}.
Then, any standard optimization algorithm (\emph{e.g.}, interior point) provides the capacity over the variable space corresponding to the input covariance matrix.
A general closed-form solution appears to be out of question, except in the case when specific conditions are satisfied, as also evidences in~\cite{tuni14} for the PAP-only case.
We derive specific conditions for the existence of a closed-form solution in the joint TP and PAP case.

Since we resort to numerical algorithms for the derivation of the channel capacity, a metric of interest is complexity.
We note that the proposed algorithm for the full-rank case reduces dramatically complexity because
$i)$ the optimization is based on a diagonal matrix determined by $\nT$ real parameters instead of the $\mathsf{n_T^\mathrm{2}}$ real parameters characterizing the input covariance matrix; and
$ii)$ positive semidefiniteness of the input covariance matrix is automatically enforced.

We also deal extensively with the \emph{singular case} (when the channel matrix rank is lower than the number of transmit antennas).
It is worth mentioning that this case was not addressed in~\cite{vu11} for the PAP-only case though some progress was made in an unpublished work by the same author~\cite{vuarxiv}.
The singular case was addressed, only in the MISO case, in~\cite{cao16,loyka17}.

The remainder of the paper is organized as follows.
Section~\ref{ch.sec} introduces the channel model along with the definition of TP and PAP constraints.
A basic form of the general optimization problem based on real variables is reported in \eqref{basic.opt.problem}.
Section~\ref{opt.sec} derives the equivalent optimization problems in the full-rank and singular cases (Theorems \ref{opt.full.th} and \ref{opt.singular.th}, respectively).
Here, \emph{full-rank} means that the channel matrix rank is equal to the number of transmit antennas and \emph{singular} corresponds to the opposite condition.
Explicit conditions for the existence of a closed-form solution are given in Corollary \ref{special.th}.
A general complexity measure defined in terms of the number of equivalent real variables required by the optimization problem is provided in Corollary \ref{complexity.th}.
The rank of the capacity-achieving covariance matrix is shown to be smaller than the channel matrix rank in Corollary \ref{Q.rank.th}.
Finally, Theorem \ref{unit.rank.singular.th} provides an explicit solution for the joint TP and PAP case applicable to the case of unit rank channel matrix.
This theorem extends the results from \cite{loyka17} applicable to the MISO case.
Algorithm~\ref{wf.algo} describes how to solve the optimization problem in this case and obtain the corresponding capacity-achieving input covariance matrix.
Section \ref{numerical.sec} provides numerical results to illustrate the capacity evaluation.
First, to validate the methods, numerical results are reported to reproduce results from \cite{vu11} and \cite{loyka17}.
Next, a few MIMO scenarios are presented to illustrate the applicability of the optimization proposed, in particular concerning the PAP constraint increase required to attain the water-filling capacity corresponding to the TP-only constraint.
Then, complexity evaluation is reported by considering a large number of channel matrix instances and measuring the complexity of the proposed form of the optimization problem~\eqref{opt.full.problem} against the basic general form~\ref{basic.opt.problem}.
Concluding remarks are reported in Section \ref{conclusions.sec}.

\subsection{Notation}

We denote matrices by uppercase boldface letters (\emph{e.g.}, $\Am$) and column vectors by lowercase boldface letters (\emph{e.g.}, $\av$).
We denote the entry at the $i$th row and $j$th column of $\Am$ as $(\Am)_{ij}$.
We denote by $\av_i$ and $\Am_i$ the $i$th column and the $i$th row of the matrix $\Am$, respectively, unless otherwise explicitly stated.
We denote by $\0_{m\times n}$ an all-zero matrix of dimensions $m\times n$.
We denote by $\Id_n$ the $n\times n$ identity matrix.
Matrix inequalities imply that the matrices involved are Hermitian and follow the standard definitions from~\cite{horn}:
$\Am>\Bm\Leftrightarrow(\Am-\Bm)$ is positive definite;
$\Am\ge\Bm\Leftrightarrow(\Am-\Bm)$ is positive semidefinite.
The Frobenius norm of a matrix $\Am$ is denoted by $\|\Am\|$ (similarly for a vector).
Notation $\diag(a_1,\dots,a_n)$ denotes a diagonal matrix whose elements on the main diagonal are $a_1,\dots,a_n$;
$\diag(\av)$ denotes a diagonal matrix whose elements on the main diagonal are taken from the vector $\av$;
$\diag(\Am)$ denotes a diagonal matrix whose elements are taken from the diagonal elements of the matrix $\Am$.
$\U(\Am)$ and $\L(\Am)$ are the upper and lower triangular matrices corresponding to $\Am$, i.e.,
$(\U(\Am))_{ij}=(\Am)_{ij}$ if $i<j$ and $0$ otherwise and $(\L(\Am))_{ij}=(\Am)_{ij}$ if $i>j$ and $0$ otherwise.
We denote the minimum and maximum eigenvalues of a Hermitian matrix $\Am$ as $\lambda_\mathsf{min}(\Am)$ and $\lambda_\mathsf{max}(\Am)$, respectively.
We denote $(x)_+\triangleq\max(0,x)$ and extend this definition to diagonal matrices by applying the scalar definition to each diagonal entry.
We also define $(\Am)_+$ for Hermitian matrices $\Am$ in the following way: if $\Am$ has the unitary factorization $\Am=\Um\Dm\Um^\H$ for some unitary matrix $\Um$ and diagonal matrix $\Dm$, then $(\Am)_+\triangleq\Um(\Dm)_+\Um^\H$.

\section{Channel Model}\label{ch.sec}

We consider an $\nT\times\nR$ MIMO channel with $\nT$ transmit and $\nR$ receive antennas, characterized by the standard channel equation
$$\yv=\Hm\xv+\zv$$
where $\Hm$ is the $\nR\times\nT$ channel matrix, $\xv$ is the transmitted symbol vector, $\zv$ is the received noise sample vector with joint circularly symmetric complex Gaussian distribution $\zv\sim\CN(0,\Id_\nR)$, and $\yv$ is the received signal sample vector.

The channel matrix is assumed to be known exactly at both the transmitter and the receiver endpoints (i.e., we have perfect CSIT/CSIR).
Every column of $\Hm$ contains at least one nonzero element and hence has positive norm (otherwise the corresponding transmitted signal would be useless).

In accordance with the previous assumptions, the mutual information corresponding to an input covariance matrix $\Qm$ is given by~\cite{cover}
$$I(\xv;\yv)=\log\det(\Id_\nR+\Hm\Qm\Hm^\H)$$
when $\xv\sim\CN(\0,\Qm)$,
and represents the maximum achievable rate over the MIMO channel under the input covariance constraint.

Hereafter, we consider the joint TP and PAP constrained MIMO channel capacity
\begin{equation}\label{basic.opt.problem}
  C = \max_{\Qm:\Qm\ge\0,\tr(\Qm)\le P_\mathsf{tot},\diag(\Qm)\le\Pm}\log\det(\Id_\nR+\Hm\Qm\Hm^\H)
\end{equation}
where $P_\mathsf{tot}$ characterizes the TP constraint, i.e., the maximum average transmitted power from all the antennas, which is represented by the trace of the input covariance matrix $\Qm$.
Moreover, $\Pm$ is a diagonal matrix containing the PAP upper bounds to the average power transmitted by every antenna.
If $\Pm=\diag(P_1,\dots,P_\nT)$, the constraints are given by $(\Qm)_{ii}\le P_i,i=1,\dots,\nT$, individually, or by $\diag(\Qm)\le\Pm$ in compact matrix form.
We shall assume that
\begin{align}\label{P.ineq}
  \sum_{i=1}^{\nT}P_i\ge P_\mathsf{tot}.
\end{align}
because otherwise the TP constraint would be unattainable.

The function $\log\det(\Id_\nR+\Hm\Qm\Hm^\H)$ is strictly concave over the convex set of positive semidefinite matrices $\Qm\ge\0$~\cite{cover}.
Since also the TP and PAP constraints lead to convex sets, the optimization problem is convex and can be solved by solving the KKT equations.
Moreover, since the objective function is strictly concave and the optimization problem is convex, its solution is global and unique~\cite{boyd}.

\begin{remark}
Optimization problem \eqref{basic.opt.problem} can be solved directly by using one of the several optimization algorithms proposed in the literature for nonlinear convex optimization (\emph{e.g.}, interior-point, sequential quadratic programming, active-set optimization algorithms)~\cite{bonnans}.
Since the matrix $\Qm$ is Hermitian and may have complex elements, one may consider the following equivalent \emph{real} optimization problem:
\begin{subequations}\label{X.opt.problem}
\begin{align}
\min_{\Xm}\quad  & -\log\det(\Id_\nR+\Hm\Qm\Hm^\H)\\
s.t.      \quad  & \Qm=\diag(\Xm)+\U(\Xm)+\U(\Xm)^\T\non
                 & +\j[\L(\Xm)-\L(\Xm)^\T]\ge\0\\
                 & \diag(\Qm)\le\Pm=\diag(P_1,\dots,P_\nT)\\
                 & \tr(\Qm)\le P_\mathsf{tot}
\end{align}
\end{subequations}
A closed-form general solution is out of question in the general MIMO case, whereas it is feasible in the MISO case~\cite{loyka17} and in a special full-rank case~\cite{vu11,tuni14}.
Nevertheless, the KKT equations can be used to decrease the complexity of the optimization problem itself by turning it into an equivalent one with a smaller number of unknowns.
\end{remark}

\section{Solution of the Optimization Problem}\label{opt.sec}

The optimization problem corresponding to the derivation of the MIMO channel capacity under a joint TP and PAP constraint can be written in the following standard form~\cite{boyd}:
\begin{subequations}\label{opt.problem}
\begin{align}
\min_{\Qm}\quad  & -\log\det(\Id_\nR+\Hm\Qm\Hm^\H)\\
s.t.      \quad  & \Qm\ge\0\\
                 & \diag(\Qm)\le\Pm=\diag(P_1,\dots,P_\nT)\\
                 & \tr(\Qm)\le P_\mathsf{tot}
\end{align}
\end{subequations}
This optimization problem is convex because of the strict convexity of the objective function $-\log\det(\Id_\nR+\Hm\Qm\Hm^\H)$~\cite{cover} and of the convexity of the domain (intersection of convex domains).

The solution of the optimization problem \eqref{opt.problem} depends on the rank of the channel matrix, which can be classified in two different cases.
\begin{enumerate}
\item
The \emph{full-rank} case, when the channel matrix rank is equal to the number of transmit antennas.
\item
The \emph{singular} case, when the channel matrix rank is smaller than the number of transmit antennas.
\end{enumerate}
The latter case occurs whenever, though not exclusively, $\nR<\nT$, as in the MISO case.
Both cases will be addressed in the following sections.

\subsection{Full-Rank Case ($\rank(\Hm)=\nT$)}\label{opt.full.sec}

The solution of the optimization problem \eqref{opt.problem} depends on the rank of the channel matrix $\Hm$.
In the case of full rank of $\Hm$ it can be characterized by the following theorem.

\begin{theorem}[$\rank(\Hm)=\nT$]\label{opt.full.th}
The capacity-achieving input covariance matrix for a MIMO channel with a joint TP and PAP constraint is given by
\begin{equation}\label{Qopt.full.eq}
  \Qm_\mathsf{opt} = \Km^{-1}(\Km\Dmc_\mathsf{opt}\Km-\Id_\nT)_+\Km^{-1},
\end{equation}
where $\Km\triangleq(\Hm^\H\Hm)^{1/2}$ is the full-rank matrix square root of the Gramian of the channel matrix $\Hm$ and  $\Dmc_\mathsf{opt}$ is derived by the solution of the optimization problem
\begin{subequations}\label{opt.full.problem}
\begin{align}
\min_{\Dmc}\quad & -\log\det[\Id_\nT+(\Km\Dmc\Km-\Id_\nT)_+]\\
s.t.      \quad & \Dmc>0\\
                & \tr[\Km^{-1}(\Km\Dmc\Km-\Id_\nT)_+\Km^{-1}]=P_\mathsf{tot}\\
                & \diag[\Km^{-1}(\Km\Dmc\Km-\Id_\nT)_+\Km^{-1}]\le\Pm
\end{align}
\end{subequations}
\end{theorem}
\begin{IEEEproof}
See Appendix \ref{opt.full.app}
\end{IEEEproof}
This optimization problem can be solved by a standard convex optimization algorithm (\emph{e.g.,} interior point).
In order to initialize the matrix $\Dmc$ we notice that, from the proof of Theorem \ref{opt.full.th} in Appendix~\ref{opt.full.app}, since $\LamF-\Id_\nT\le(\LamF-\Id_\nT)_+$, and hence, from \eqref{Ptot.eq}, we get
$$\Dmc-\Km^{-2}\le\Qm\Rightarrow\tr(\Dmc)\le P_\mathsf{tot}+\tr(\Km^{-2}).$$
Thus, we can initialize $\Dmc$ by
$$\Dmc_0\triangleq(P_\mathsf{tot}+\alpha\tr(\Km^{-2})\Id_\nT$$
for some $\alpha<1$.
The choice of $\alpha$ affects the convergence properties of the iterative algorithm.

As noted in \cite{vu11}, and subsequently in \cite{tuni14}, for the PAP-only case, a special case occurs when it is possible to solve in closed form the optimization problem of Theorem \ref{opt.full.th}.
With the joint TP and PAP constraints we have the the following result.

\begin{corollary}[$\rank(\Hm)=\nT$]\label{special.th}
The capacity-achieving input covariance matrix for a MIMO channel with joint TP and PAP constraints is
\begin{equation}\label{Qopt.special.eq}
  \Qm_\mathsf{opt} = \frac{P_\mathsf{tot}+\tr(\Km^{-2})}{\nT}\Id_\nT-\Km^{-2},
\end{equation}
where $\Km\triangleq(\Hm^\H\Hm)^{1/2}$, if the following conditions hold:
\begin{subequations}
\begin{align}
P_\mathsf{tot}&\ge\nT\lambda_\mathsf{max}(\Km^{-2})-\tr(\Km^{-2})\\
P_\mathsf{tot}&\le\nT\min_{1\le i\le\nT}\Big\{(\Km^{-2})_{ii}+P_i\Big\}-\tr(\Km^{-2})
\end{align}
\end{subequations}
The capacity is
\begin{equation}
C=\nT\log_2\bigg(\frac{P_\mathsf{tot}+\tr(\Km^{-2})}{\nT}\bigg)+\log_2(\Km^2).
\end{equation}
\end{corollary}

\begin{IEEEproof}
Assume that $\Dmc>\Km^{-2}$ and maximize $\det(\Dmc)$ by the scaled identity matrix in the proof of Theorem \ref{opt.full.th}.
\end{IEEEproof}

\subsection{Singular Case ($\rank(\Hm)<\nT$)}\label{opt.singular.sec}

Here, we consider the case of a channel matrix with rank strictly lower than the number of columns, i.e., the number of transmit antennas $\nT$.
The following theorem characterizes the capacity-achieving covariance matrix in the singular case.

\begin{theorem}[$\rank(\Hm)<\nT$]\label{opt.singular.th}
The capacity-achieving input covariance matrix for a MIMO channel with joint TP and PAP constraints and channel matrix with rank $\nu<\nT$, the number of transmit antennas, is obtained by solving the following optimization problem.
\begin{subequations}\label{opt.singular.problem}
\begin{align}
\min_{\Dmc,\Qm}\quad & -\log\det[\Id_\nR+\Hm\Qm\Hm^\H]\\
s.t.           \quad & \Dmc>0,\Qm\ge\0\\
                     & \diag(\Qm)\le\Pm\\
                     & \tr(\Qm)\le P_\mathsf{tot}\\
                     & \VmH^\H\Qm\VmH=\LamH^{-1}\UmH^\H(\Hm\Dmc\Hm^\H-\Id_\nR)_+\UmH\LamH^{-1}
\end{align}
\end{subequations}
where we used the reduced-size SVD of the channel matrix $\Hm=\UmH\LamH\VmH^\H$, where $\LamH$ is a $\nu\times\nu$ positive definite diagonal matrix and $\UmH,\VmH$ are $\nT\times\nu$ and $\nR\times\nu$ partial unitary matrices, such that $\UmH^\H\UmH=\VmH^\H\VmH=\Id_\nu$.
\end{theorem}
\begin{IEEEproof}
See Appendix \ref{opt.singular.app}
\end{IEEEproof}

\begin{remark}
Notice that, in the singular case, the TP constraint is not necessarily met with equality because the relative argument used in the case of nonsingular $\Km$ (deriving from the inequality $\diag(\Qm)\le\Pm$) fails to hold in this case where the PAP constraints implies that $\diag(\Vm_+\Qmt_+\Vm_+^\H)\le\Pm$.
This has a major impact on the capacity, which can be substantially lower in this case because meeting the PAP constraints prevents to exploit the total available power according to the TP constraint.
\end{remark}

\begin{corollary}[Complexity]\label{complexity.th}
The complexity of the equivalent optimization problem derived in Ths.\ \ref{opt.full.th} and \ref{opt.singular.th}, in terms of cardinality of the variable space, is given by
\begin{equation}\label{complexity.eq}
N_\mathsf{var}=2(\nT-\nu)\nu+\nT.
\end{equation}
\end{corollary}

\begin{IEEEproof}
See Appendix \ref{opt.singular.app}.
\end{IEEEproof}

\begin{remark}
It is interesting to notice that the worst-case complexity of the optimization algorithm corresponds to the case of rank $\nu=\nT/2$ or the closest integer.
Compared to the fixed complexity of the basic form of the optimization problem \eqref{X.opt.problem}, the complexity of the optimization algorithm in the full-rank case is $\nT$ times lower.
When $\nu=\nT/2$, the complexity is still reduced by a factor $2$ and when $\nu=1$, the reduction is by a factor $\nT/3$, approximately.
However, in this last case, the use of Algorithm \ref{wf.algo} from Appendix \ref{unit.rank.singular.app} dramatically reduces complexity and provides an exact result without the need of numerical algorithms like interior-point.
\end{remark}

\begin{corollary}[Rank]\label{Q.rank.th}
The capacity-achieving input covariance matrix $\Qm$ for a MIMO channel with joint TP and PAP constraints and channel matrix with rank $\nu<\nT$ satisfies the following inequalities:
\begin{equation}\label{Q.rank.eq}
1\le\rank(\Qm)\le\nu.
\end{equation}
\end{corollary}

\begin{IEEEproof}
See Appendix \ref{opt.singular.app}.
\end{IEEEproof}

\subsection{Unit-Rank Singular Case}

In this section we consider a MIMO channel with channel matrix having unit rank, i.e., $\nu=1$.
The MISO channel is a special case where $\nu=\nR=1$.
More generally, a unit-rank MIMO channel has a channel matrix which can be expressed as
\begin{equation}\label{unit.rank.H.eq}
\Hm=\|\Hm\|\uv\vv^\H,
\end{equation}
where $\uv,\vv$ are unit-norm column vectors.

The special MISO case has been dealt with in the literature (see, in particular, \cite{cao16,loyka17}) where closed-form expressions have been proposed.
Our approach extends these results and our findings are summarized in the following theorem.

\begin{theorem}\label{unit.rank.singular.th}
Given a MIMO channel with unit-rank channel matrix having SVD $\Hm=\|\Hm\|\uv\vv^\H$, where $\|\uv\|=\|\vv\|=1$, the channel capacity with joint TP and PAP constraints is given by
\begin{equation}\label{unit.rank.singular.cap.eq}
C=\log\bigg(1+\|\Hm\|^2\sum_{i=1}^{\nT}|v_iq_i|^2\bigg)
\end{equation}
where $v_i,q_i$ are the $i$th elements of the vectors $\vv,\qv$, respectively, for $i=1,\dots,\nT$, and the vector $\qv$ is obtained by setting $$q_i\triangleq\sqrt{\min(\alpha|v_i|^2,P_i)}\ \e^{-\j\angle v_i},i=1,\dots,\nT,$$ where $\alpha$ is obtained by solving the equation
\begin{equation}\label{alpha.eq}
\sum_{i=1}^{\nT}\min(\alpha|v_i|^2,P_i)=P_\mathsf{tot}.
\end{equation}
The unit-rank capacity-achieving covariance matrix is
\begin{equation}\label{unit.rank.singular.Q.eq}
\Qm_\mathsf{opt}=\qv\qv^\H
\end{equation}
and the TP constraint is always attained with equality.
\end{theorem}
\begin{IEEEproof}
See Appendix \ref{unit.rank.singular.app}.
The appendix reports an algorithm for the solution of \eqref{alpha.eq}, Algorithm~\ref{wf.algo}.
\end{IEEEproof}

Theorem \ref{unit.rank.singular.th} extends the results of~\cite[Th.1,2]{loyka17} and~\cite{cao16} from the MISO case to the unit-rank MIMO case.
It obviously applies to the MISO case, as well.
Algorithm~\ref{wf.algo} from Appendix~\ref{unit.rank.singular.app} is more easily implementable than the derivations in~\cite{cao16,loyka17}.
It is also closely related to the algorithm used to solve the water-filling equation.

\section{Numerical Results}\label{numerical.sec}

In this section we report some numerical results obtained by applying the proposed optimization algorithm implemented in Matlab.

\subsection{Validation}

Fig.~\ref{f1.fig} validates the results of the proposed algorithm against those reported in~\cite{vu11}.
The diagrams show the capacity of a $2\times2$ MIMO channel with the channel matrix defined in~\cite[eq.\ (26)]{vu11} in the following cases:
{\sf\small PAP}: PAP constraint with $P_1$ from the abscissa and $P_2=1-P_1$;
{\sf\small TP}: TP constraint with $P_\mathsf{tot}=1$ (water-filling solution);
{\sf\small MC}: $\Qm=\diag(P_1,1-P_1)$.
It can be seen that the curves coincide exactly with those reported in~\cite{vu11}.

\insertfig{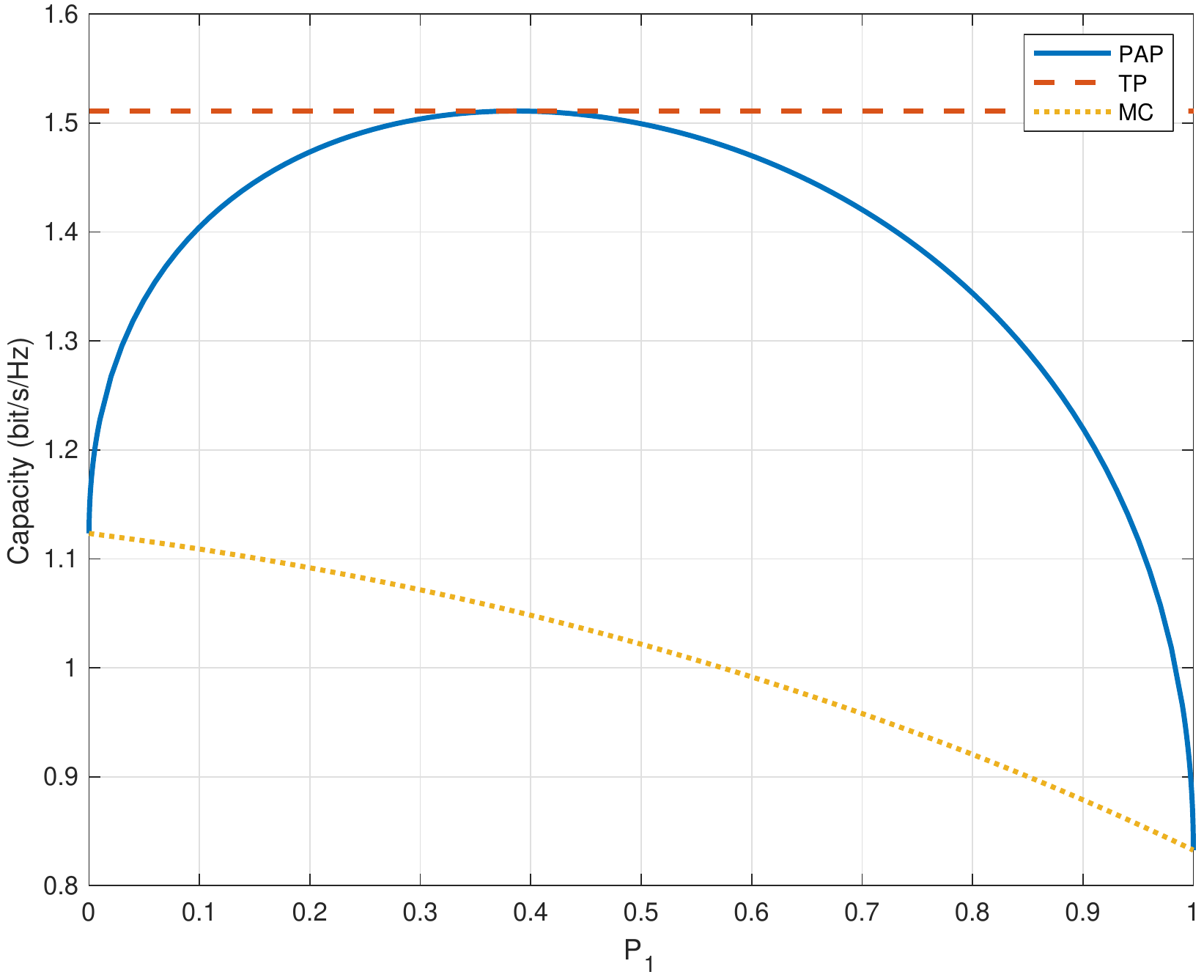}{Capacity of a $2\times2$ MIMO channel with channel matrix defined in~\cite[eq.\ (26)]{vu11} in the following cases:
{\sf\scriptsize PAP}: PAP constraint with $P_1$ from the abscissa and $P_2=1-P_1$;
{\sf\scriptsize TP}: TP constraint with $P_\mathsf{tot}=1$;
{\sf\scriptsize MC}: $\Qm=\diag(P_1,1-P_1)$.}

Figs.\ \ref{f2a.fig}--\ref{f2b.fig} validate the results from Algorithm \ref{wf.algo} against those of \cite[Figs.1-2]{loyka17}.
The former reports the capacity versus the TP constraint $P_\mathsf{tot}$ and the latter reports the $\nT=4$ normalized amplitudes $[(\Qm)_{ii}/P^*]^{1/2}$ for $i=1,\dots,\nT$ of the capacity-achieving covariance matrix after defining $P^*\triangleq\min(P_\mathsf{tot},\tr(\Pm))$.
It can be seen that the curves coincide exactly with those reported in~\cite{loyka17}.

\insertfig{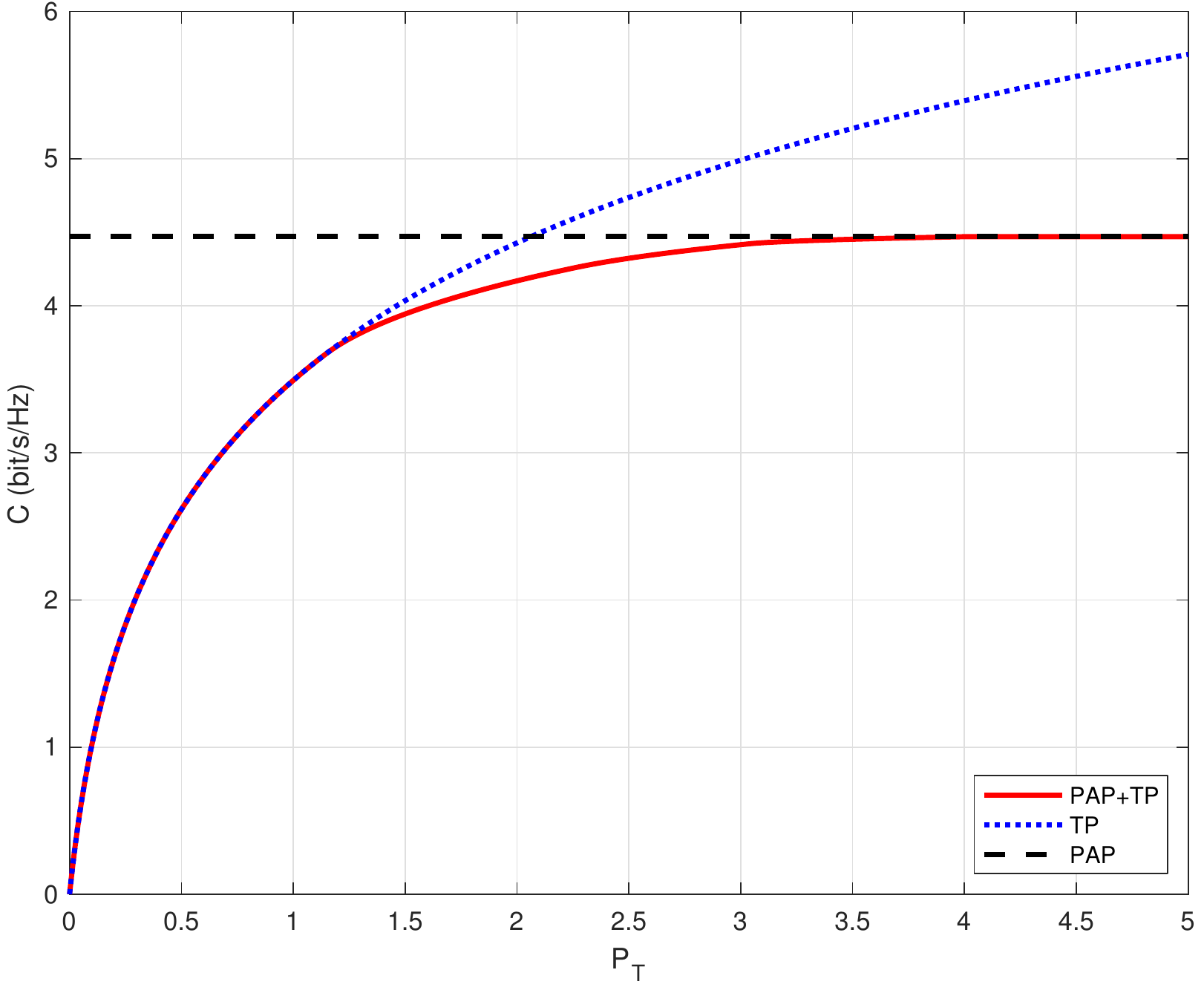}{Capacity of a $4\times1$ MISO channel with matrix from~\cite[Fig.\ 1]{loyka17}.}
\insertfig{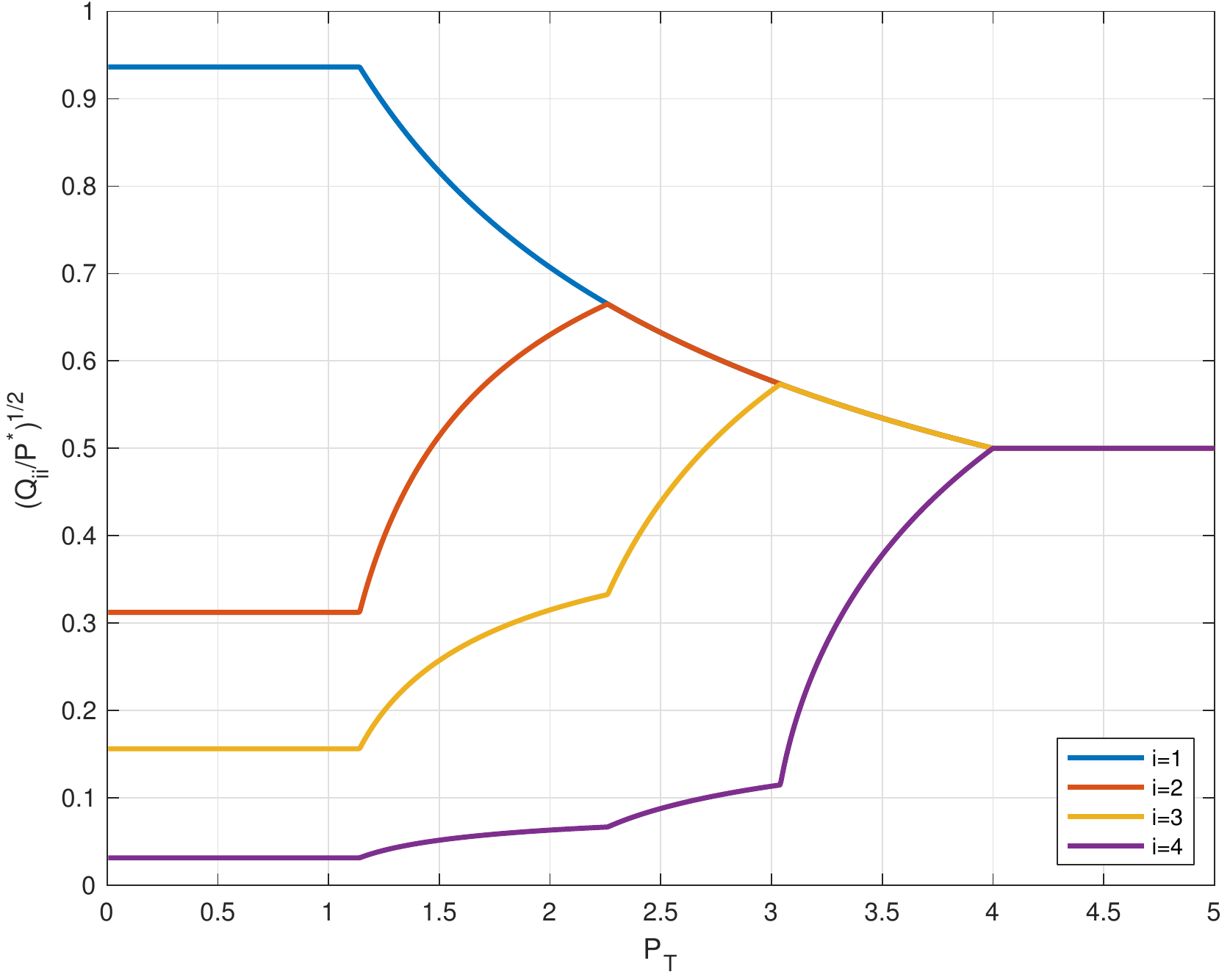}{Amplitude distribution from the capacity-achieving covariance matrix normalized with respect to $P^*\triangleq\min(P_\mathsf{tot},\tr(\Pm))$ with channel matrix as in Fig.\ \ref{f2a.fig}.}

\subsection{Applications}

Next, we illustrate the application of the algorithms proposed in Theorem \ref{opt.full.th} and \ref{opt.singular.th} by considering the $3\times4$ and $3\times2$ channel matrices obtained in Matlab by the following commands:
\begin{itemize}
\item \verb|rng(1);H1=randn(4,3)+1i*randn(4,3);|
\item \verb|rng(1);H2=randn(2,3)+1i*randn(2,3);|
\end{itemize}
The matrices obtained are given in eqs.\ \eqref{H1.eq} and \eqref{H2.eq}.
\begin{figure*}
\begin{equation}\label{H1.eq}
\Hm_\mathsf{3\times4}=\begin{pmatrix}
  -0.6490 - 1.5094\j & -0.8456 - 1.9654\j & -0.1969 - 0.2752\j \\
   1.1812 + 0.8759\j & -0.5727 - 1.2701\j &  0.5864 + 0.6037\j \\
  -0.7585 - 0.2428\j & -0.5587 + 1.1752\j & -0.8519 + 1.7813\j \\
  -1.1096 + 0.1668\j &  0.1784 + 2.0292\j &  0.8003 + 1.7737\j
\end{pmatrix}
\end{equation}
\begin{equation}\label{H2.eq}
\Hm_\mathsf{3\times2}=\begin{pmatrix}
  -0.6490 - 0.5587\j & -0.7585 - 0.1969\j & -0.8456 - 0.8519\j \\
   1.1812 + 0.1784\j & -1.1096 + 0.5864\j & -0.5727 + 0.8003\j
\end{pmatrix}
\end{equation}
\hrulefill
\end{figure*}
$\Hm_1$ corresponds to an instance of full-rank MIMO since $\rank(\Hm_1)=\nT=3$ and
$\Hm_1$ corresponds to an instance of singular MIMO since $\rank(\Hm_1)=2<\nT=3$.
In both cases we assume $\Pm=\diag(0.1,0.1,1)$ and plot the capacity with respect to $P_\mathsf{tot}$.
The results obtained are illustrated in Fig.\ \ref{f3.fig}.
We can see that the rank of the capacity-achieving covariance matrices increases with the TP constraint up to the maximum predicted by Corollary \ref{Q.rank.th}.
Also, the capacity with the joint constraint saturates when the TP constraint exceeds the sum of the PAP constraints, which is equal to $1.2$ in this example.

\insertfig{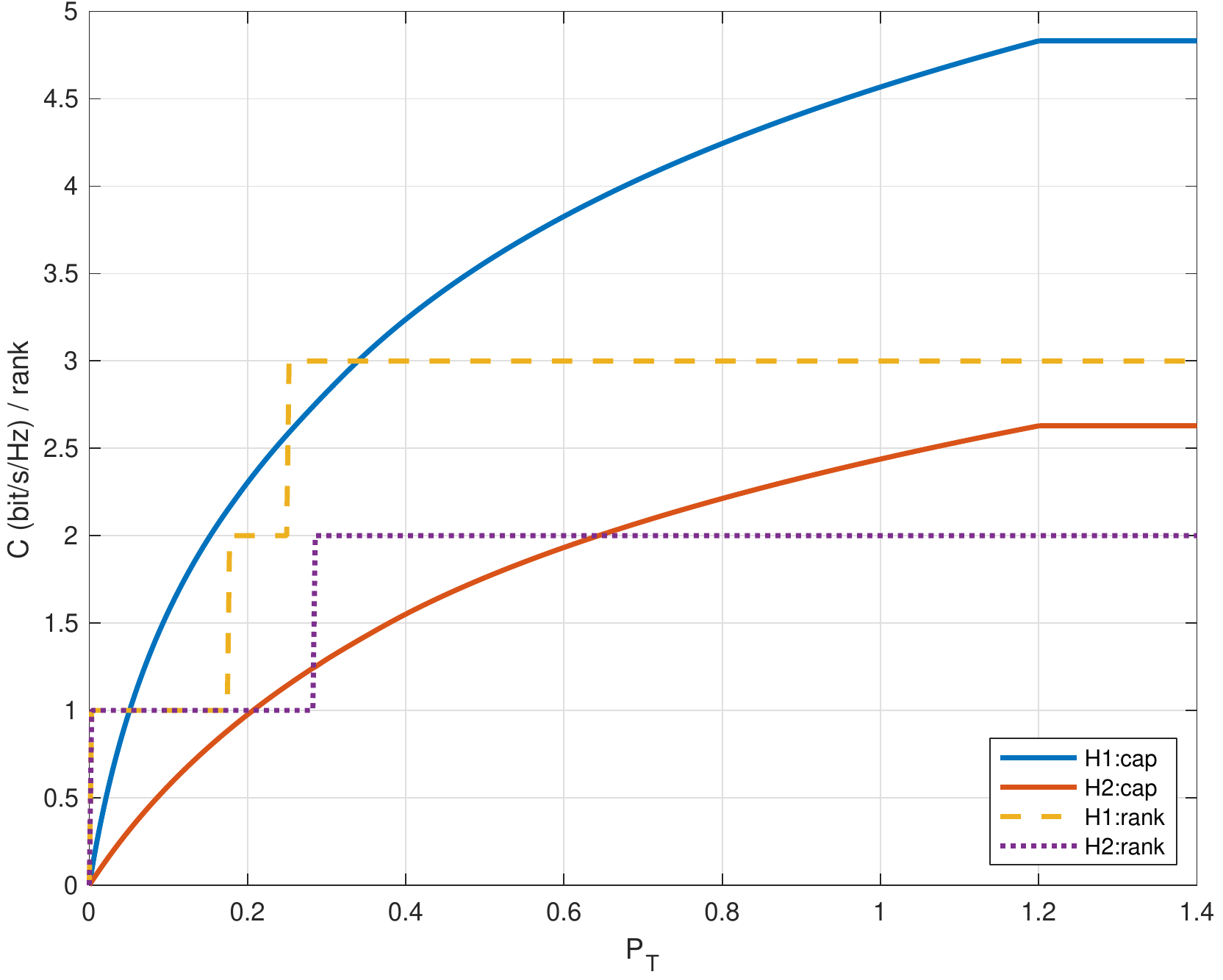}{Capacity of the $3\times4$ and $2\times4$ MIMO channels corresponding to the channel matrices reported in \eqref{H1.eq} and \eqref{H2.eq} with fixed PAP constraints given by $\Pm=\diag(0.1,0.1,1)$ and variable TP constraint in abscissa.
The curves labeled by {\sf\scriptsize H1:cap} and {\sf\scriptsize H2:cap} report the capacity corresponding to the channel matrices $\Hm_1$ and $\Hm_2$, respectively.
The piecewise horizontal lines labeled by {\sf\scriptsize H1:rank} and {\sf\scriptsize H2:rank} report the rank of the capacity-achieving covariance matrices corresponding to the channel matrices reported in \eqref{H1.eq} and \eqref{H2.eq}, respectively.
}

Another illustration of the joint TP and PAP constrained capacity calculation is given in Fig.\ \ref{f5a.fig}.
The figure considers a $3\times3$ and a $4\times4$ MIMO channel with channel matrices given in \eqref{MIMO3x3.eq} and \eqref{MIMO4x4.eq}, respectively.
The TP constraint is $P_\mathsf{tot}=3$ and $4$, respectively, and the PAP constraint is in the abscissa, constant for every antenna.
It can be seen that equal power allocation achieves a lower capacity than variable power allocation based on water-filling and the TP constraint, as expected.
Increasing the PAP constraint increases the capacity up to the water-filling limit imposed by the TP constraint, as illustrated by the curves of Fig.\ \ref{f5a.fig}.
Similarly, Figs.\ \ref{f5b.fig} and \ref{f5c.fig} illustrate the same scenario with TP constraint reduced by a factor $10$ and $100$, respectively.
The figures show that, as the TP constraint decreases, the PAP constraint must increase more and more in order to achieve the maximum capacity corresponding to a TP-only constrained system.
This is a consequence of the fact that the water-filling power distribution increases its dynamic range as the TP constraint decreases.

\begin{figure*}
\begin{equation}\label{MIMO3x3.eq}
\Hm_\mathsf{3\times3}=\begin{pmatrix}
   0.1038 - 0.0877\j & -0.4125 - 1.6836\j &  1.9318 + 0.2237\j\\
  -0.0410 - 0.0299\j & -0.5255 - 0.5724\j & -0.7740 - 1.4445\j\\
   0.0074 + 0.1378\j & -0.6510 + 0.4731\j &  1.4142 + 0.8371\j
\end{pmatrix}
\end{equation}
\begin{equation}\label{MIMO4x4.eq}
\Hm_\mathsf{4\times4}=\begin{pmatrix}
   0.3576 + 0.1914\j &  0.0614 + 0.1692\j & -0.7338 + 1.0030\j & -0.0943 - 0.3671\j\\
   0.7547 + 0.1277\j &  0.5032 - 0.2920\j & -0.4087 - 1.9283\j &  0.3410 - 0.0309\j\\
  -0.4020 + 1.1647\j &  0.1404 + 0.3537\j &  1.6532 + 3.0492\j & -0.4247 - 0.0599\j\\
   0.7130 - 2.0329\j &  0.4724 + 0.5394\j & -0.2910 + 0.7012\j & -0.7934 + 0.2927\j
\end{pmatrix}
\end{equation}
\hrulefill
\end{figure*}

\insertfig{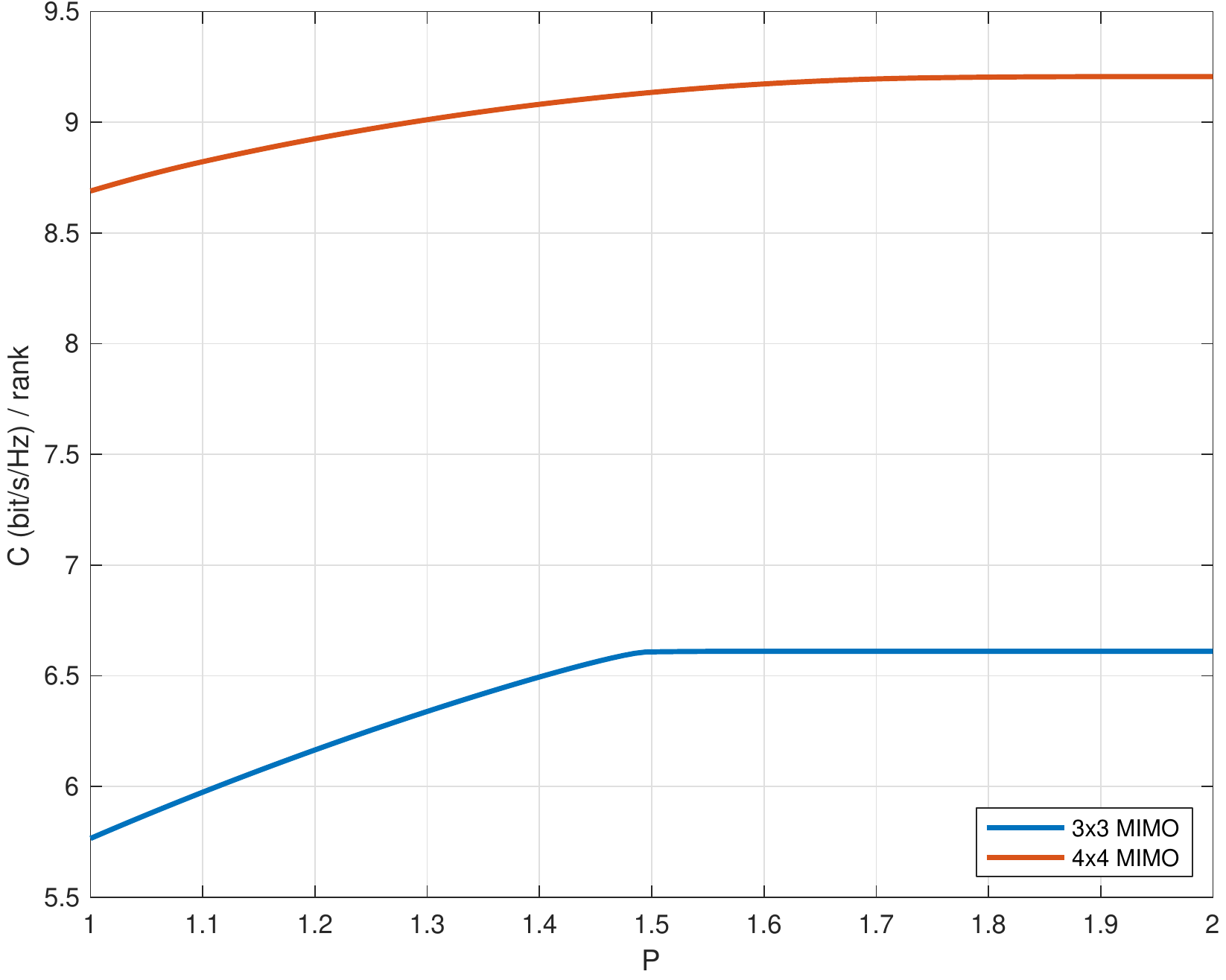}{Capacity of the $3\times3$ and $4\times4$ MIMO channels, described by the channel matrices in \eqref{MIMO3x3.eq} and \eqref{MIMO4x4.eq}, respectively, versus a constant PAP constraint $P$ in abscissa and a TP constraint $P_\mathsf{tot}=3$ and $4$, respectively.}
\insertfig{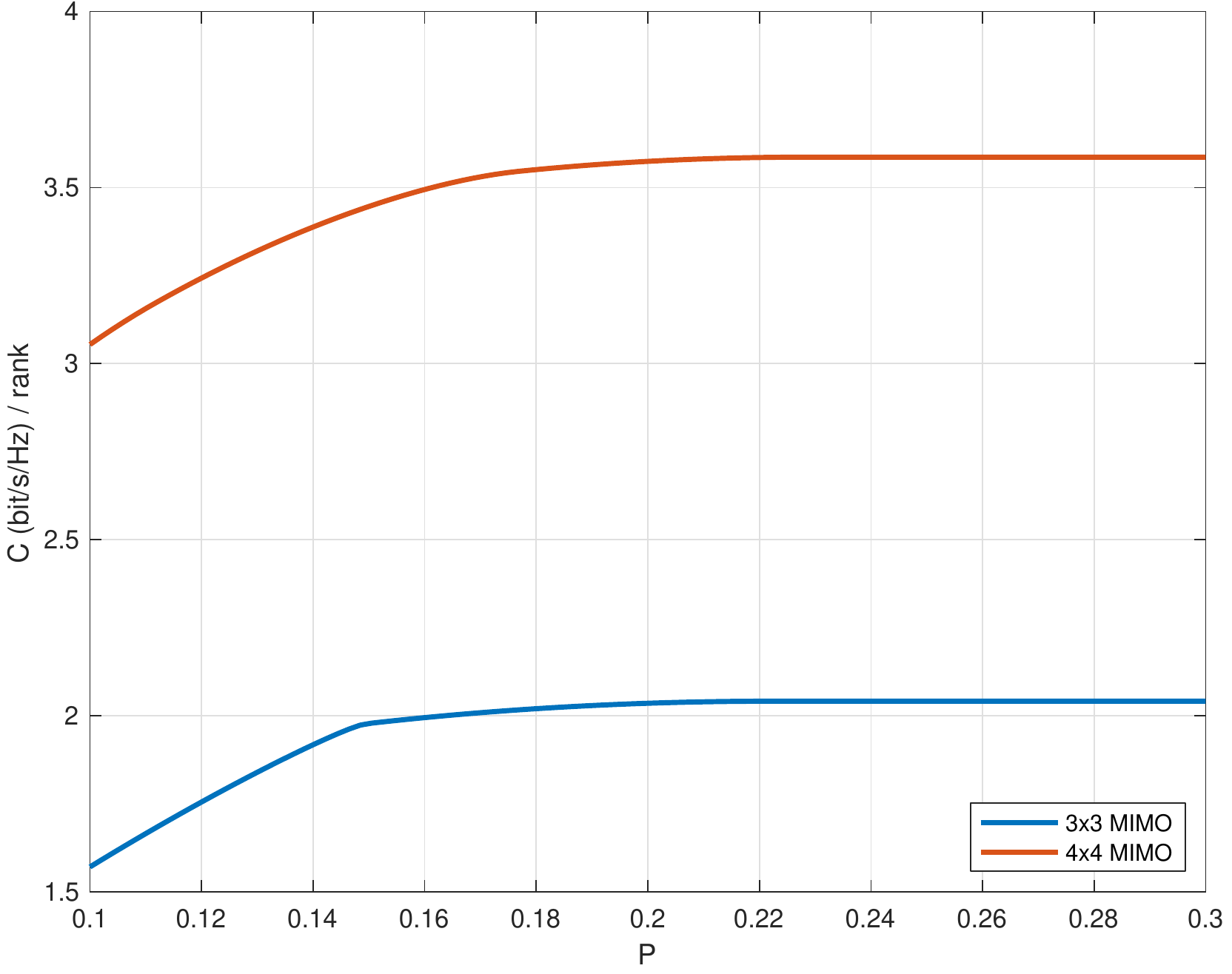}{Same as Fig.\ \ref{f5a.fig}, but with TP constraint $P_\mathsf{tot}=0.3$ and $0.4$, respectively.}
\insertfig{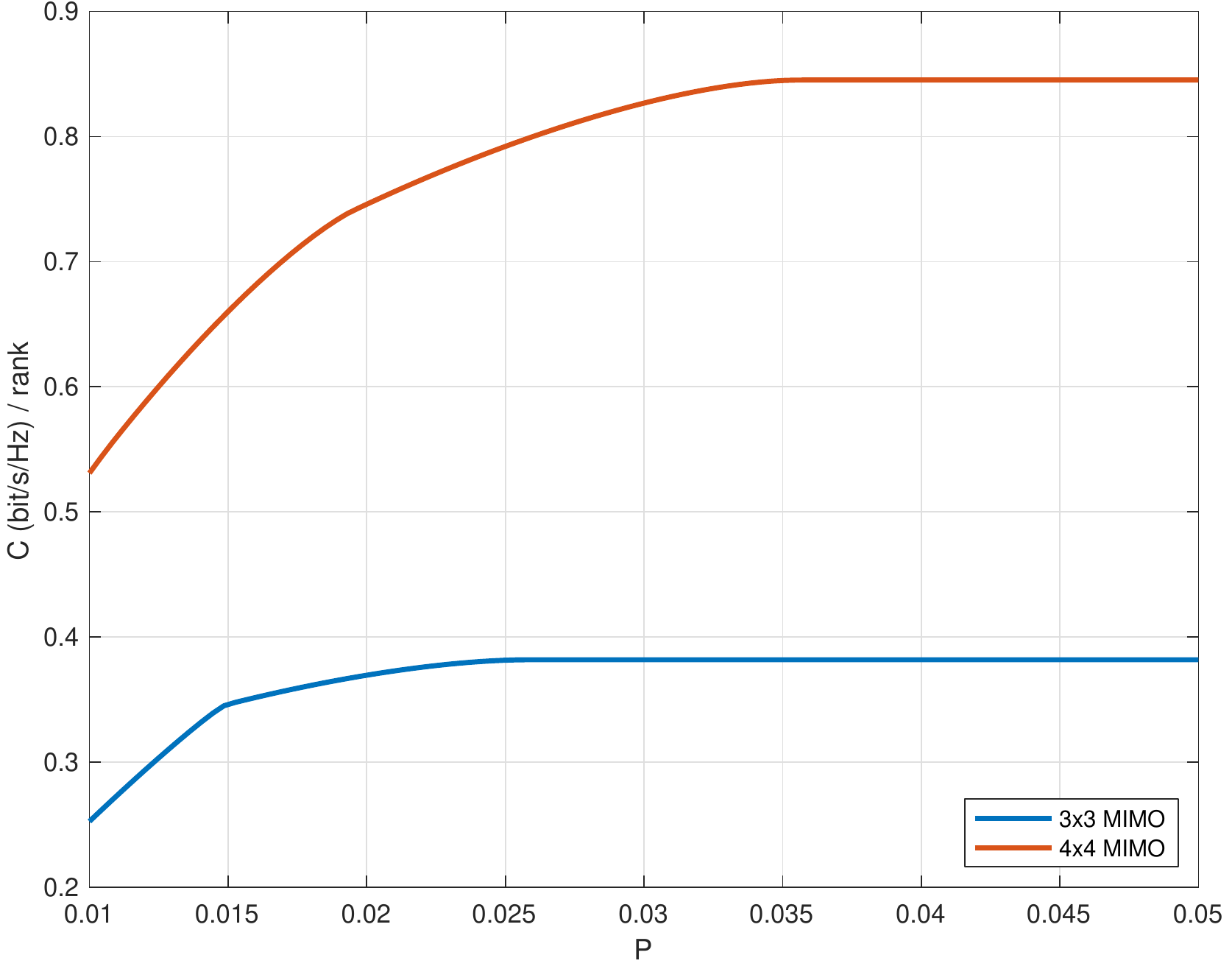}{Same as Fig.\ \ref{f5a.fig}, but with TP constraint $P_\mathsf{tot}=0.03$ and $0.04$, respectively.}

\subsection{Complexity}

The time complexity of numerical optimization is illustrated in Fig.\ \ref{f4.fig}.
The points report the average CPU time to calculate the capacity of an $n\times n$ MIMO channel with iid Rayleigh distributed channel gains.
The PAP constraint is set to $$\Pm=\diag(1,\underbrace{0.1,\dots,0.1}_{n-1})$$ and the TP constraint is set to $P_\mathsf{tot}=1$.
We can see that the time complexity growth linearly with $\nT$ and the proposed optimization algorithm based on  \eqref{opt.full.problem}, which is a consequence of the fact that the variable space is $\nT$-dimensional.
The basic optimization algorithm based on problem \eqref{basic.opt.problem} is far more expensive in terms of cpu-time complexity.
The simulation results show a more than cubic growth with $\nT$, more than the increased dimension of the variable space, which is $\mathsf{n_T^\mathrm{2}}$ in this case.
This excessive growth can be partly explained by noting that the basic algorithm requires a positive definiteness test in the constraint equations, whose complexity is cubic in the matrix size.
Needless to say, the analytic complexity evaluation of optimization algorithms is a hard topic and we limit ourselves to make a comparison based on simulation results.
In the unit-rank case, Algorithm \ref{wf.algo} attains the minimum time complexity, several orders of magnitude lower than the complexity of solving the optimization problems~\eqref{basic.opt.problem} and~\eqref{opt.singular.problem}.

\insertfig{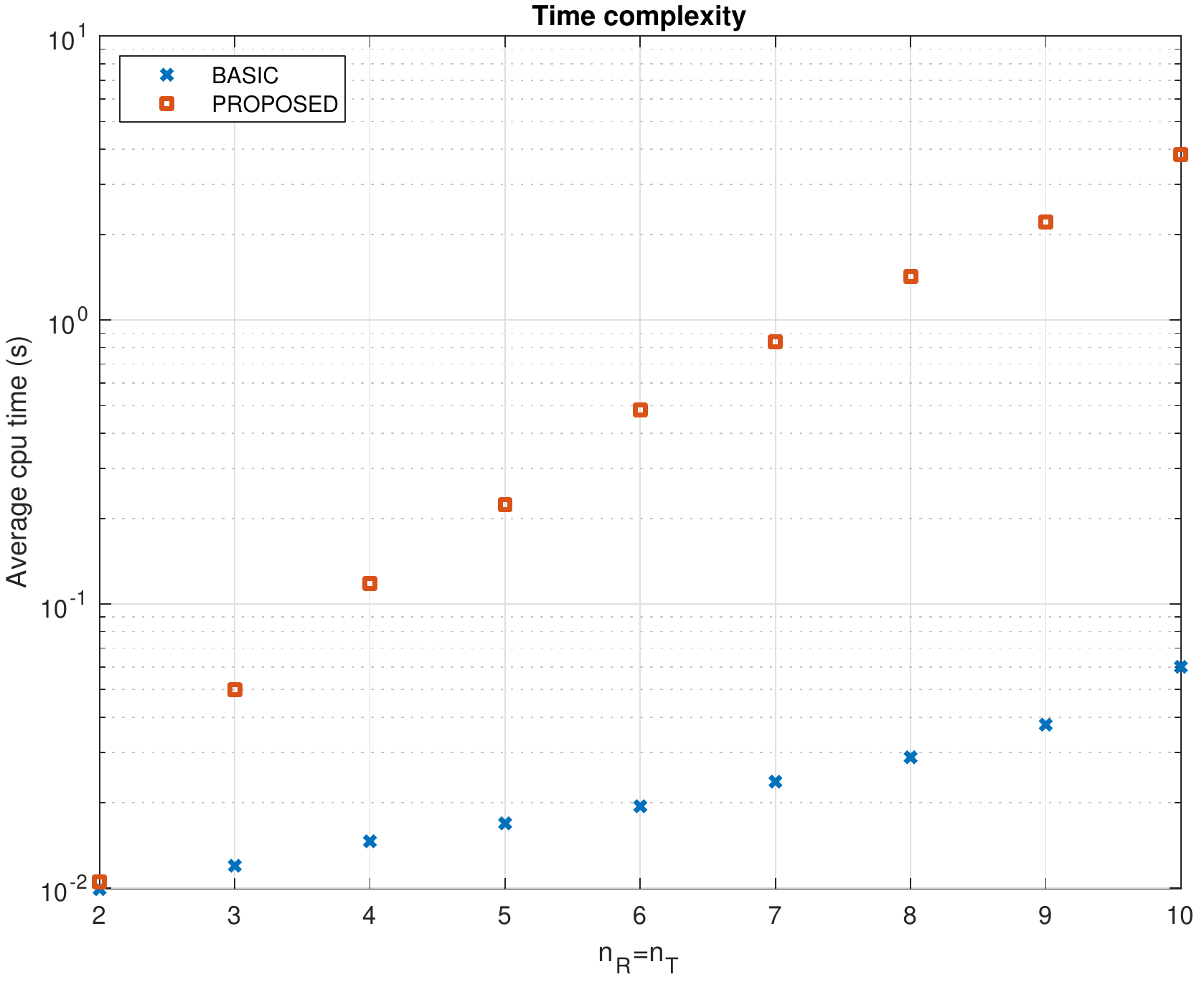}{Average cpu time versus number of antennas based on the optimization problems \eqref{basic.opt.problem} (BASIC) and \eqref{opt.full.problem} (PROPOSED).
Random iid Rayleigh $n\times n$ MIMO channel.}

\section{Conclusions}\label{conclusions.sec}

In this work we addressed the derivation of the capacity of a MIMO channel under joint total and per-antenna average power constraints.
We have considered separately the \emph{full-rank} and \emph{singular} cases, which consist, respectively, of having the channel matrix rank equal to or lower than the number of transmit antennas.

Closed-form results have been provided in a special full-rank case (Corollary~\ref{special.th}) and for the general unit-rank MIMO case (Theorem~\ref{unit.rank.singular.th}).
The unit-rank case is supported by an iterative algorithm similar to the \emph{water-filling} algorithm leading to the unit-rank capacity achieving covariance matrix (Algorithm~\ref{wf.algo}).

When the channel rank is greater than one, numerical optimization seems to be the only way to find the channel capacity.
Then, we focused on the complexity of numerical optimization, identified by the number of equivaent real variables in the optimization problem.
First, we noticed that the general case can be always solved by an optimization problem depending on $\mathsf{n_T^\mathrm{2}}$ real variables, i.e., problem~\eqref{basic.opt.problem}.
Then, we derived equivalent optimization problems with a smaller number of equivalent real parameters for the full-rank case (Theorem~\ref{opt.full.th}) and for the singular case (Theorem~\ref{opt.singular.th}).
The total number of equivalent real parameters is given by Corollary~\ref{complexity.th}.
The range of possible values of the capacity-achieving input covariance matrix rank is derived in Corollary~\ref{Q.rank.th}.

Finally, numerical results are reported
$i)$ to validate the optimization algorithms against existing literature results;
$ii)$ to illustrate the application of the proposed methods to practical MIMO scenarios where the per-antenna constraint is of major interest; and
$iii)$ to evaluate the time complexity of the proposed algorithms.

\appendices
\section{Proof of Theorem \ref{opt.full.th}}\label{opt.full.app}

\begin{IEEEproof}
The optimization problem leading to the capacity considered in Theorem \ref{opt.full.th} leads to the Lagrangian function given by
\begin{align*}
\L(\Qm)=&
-\log\det(\Id_\nR+\Hm\Qm\Hm^\H)+\lambda[\tr(\Qm)-P_\mathsf{tot}]\\
&+\tr(\Lam(\Qm-\Pm))-\tr(\Mm\Qm).
\end{align*}
The KKT equations can be written as follows:
\begin{subequations}\label{kkt.eq}
\begin{align}
\Hm^\H(\Id_\nR+\Hm\Qm\Hm^\H)^{-1}\Hm&=\lambda\Id_\nT+\Lam-\Mm\\
\lambda&\ge0\\
\text{Diagonal}\ \Lam&\ge\0\\
\text{Hermitian}\ \Mm,\Qm&\ge\0\\
\Mm\Qm&=\0\\
\Lam(\diag(\Qm)-\Pm))&=\0\\
\lambda[\tr(\Qm)-P_\mathsf{tot}]&=0\\
\diag(\Qm)&\le\Pm\\
\tr(\Qm)&\le P_\mathsf{tot}
\end{align}
\end{subequations}
The diagonal matrix $\lambda\Id_\nT+\Lam$ is positive definite since
\begin{align*}
(\lambda\Id_\nT+\Lam)_{ii}
&=\lambda+\lambda_i\\
&=\hv_i^\H(\Id_\nR+\Hm\Qm\Hm^\H)^{-1}\hv_i+(\Mm)_{ii}\\
&\ge\hv_i^\H(\Id_\nR+\Hm\Qm\Hm^\H)^{-1}\hv_i\\
&>0.
\end{align*}
The inequalities depend on the fact that $\Mm\ge\0$ (hence $(\Mm)_{ii}\ge0$) and that $\Id_\nR+\Hm\Qm\Hm^\H>\0$ (since $\Qm\ge\0$).
The last inequality holds unless $\hv_i=\0$, which would imply no signal transmission from the $i$-th antenna.
Next, we define the diagonal matrix
\begin{equation}
  \Dm\triangleq\lambda\Id_\nT+\Lam>0.
\end{equation}
From the KKT equations \eqref{kkt.eq} we get the following equations:
\begin{subequations}
\begin{align}
\Hm^\H(\Id_\nR+\Hm\Qm\Hm^\H)^{-1}\Hm\Qm&=\Dm\Qm\\
\text{Diagonal}\ \Dm&>\0\\
\diag(\Qm)&\le\Pm\\
\tr(\Qm)&\le P_\mathsf{tot}\label{Ptot.ineq}
\end{align}
\end{subequations}
Now, the constraint \eqref{Ptot.ineq} must be met with equality.
Assume, on the contrary, that the optimum covariance matrix $\Qm_\mathsf{opt}$ satisfies the inequality $\tr(\Qm_\mathsf{opt})<P_\mathsf{tot}$.
Then, we consider a covariance matrix $\Qm(\alpha)$ defined by
$$(\Qm(\alpha))_{ij}\triangleq\bigg(\frac{P_i}{(\Qm_\mathsf{opt})_{ii}}\bigg)^{\alpha/2}(\Qm)_{ij}
\bigg(\frac{P_j}{(\Qm_\mathsf{opt})_{jj}}\bigg)^{\alpha/2}.$$
It is plain to see that $\Qm(\alpha)\ge\0$.
Moreover, $\Qm(0)\equiv\Qm_\mathsf{opt}$ and $\Qm(1)>\Qm_\mathsf{opt}$ since at least for one index $i$ with $1\le i\le\nT$ we have $(\Qm_\mathsf{opt})_{ii}<P_i$ (otherwise, if $(\Qm_\mathsf{opt})_{ii}=P_i$ for every $i=1,\dots,\nT$, it would be $\tr(\Qm_\mathsf{opt})=\sum_{i=1}^{\nT}(\Qm)_{ii}=\sum_{i=1}^{\nT}P_i\ge P_\mathsf{opt}$ from \eqref{P.ineq}).
Next, we define
$$\tau(\alpha)\triangleq\tr[\Qm(\alpha)]=\sum_{i=1}^{\nT}\bigg(\frac{P_i}{(\Qm_\mathsf{opt})_{ii}}\bigg)^{\alpha}(\Qm_\mathsf{opt})_{ii}.$$
We can see that $\tau(\alpha)$ is a continuous monotonically increasing real function of $\alpha$ for $\alpha\ge0$.
Moreover, $\tau(0)=\tr(\Qm_\mathsf{opt})<P_\mathsf{tot}$ by assumption, and
$$\tau(1)=\sum_{i=1}^{\nT}P_i>P_\mathsf{tot}$$
by inequality \eqref{P.ineq}.
Hence, there exists $\hat{\alpha}\in(0,1)$ such that $\tau(\hat{\alpha})=P_\mathsf{tot}$ and $\Qm(\hat{\alpha})>\Qm_\mathsf{opt}$ satisfying the total power constraint with equality and attaining a greater mutual information $\log\det(\Id_\nR+\Hm\Qm(\hat{\alpha})\Hm^\H)>\log\det(\Id_\nR+\Hm\Qm_\mathsf{opt}\Hm^\H)$, contrary to the assumed optimality of $\Qm_\mathsf{opt}$.

As a consequence, the optimization problem to be solved to find the capacity is given by:
\begin{subequations}\label{opt.xxx.problem}
\begin{align}
\Hm^\H(\Id_\nR+\Hm\Qm\Hm^\H)^{-1}\Hm\Qm&=\Dm\Qm \label{xxx.eq}\\
\text{Diagonal}\ \Dm&>\0\\
\diag(\Qm)&\le\Pm\\
\tr(\Qm)&=P_\mathsf{tot} \label{Ptot.eq}
\end{align}
\end{subequations}
where $(\Qm)_{ii}<P_i$ for at least one index $i\in\{1,\dots,\nT\}$.

Now we find an equivalent expression of \eqref{xxx.eq}:
\begin{subequations}
\begin{align}
\eqref{xxx.eq}\stackrel{(a)}{\Rightarrow}
&\Hm\Dm^{-1}\Hm^\H(\Id_\nR+\Hm\Qm\Hm^\H)^{-1}\Hm\Qm\Hm^\H=\Hm\Qm\Hm^\H\label{yyy1.eq}\\
\stackrel{(b)}{\Rightarrow}
&\Hm\Dm^{-1}\Hm^\H\Hm\Qm\Hm^\H=\Hm\Qm\Hm^\H(\Id_\nR+\Hm\Qm\Hm^\H)\label{yyy2.eq}\\
\stackrel{(c)}{\Rightarrow}
&\Fm\Rm=\Rm+\Rm^2 \label{yyy3.eq}
\end{align}
\end{subequations}
where we
$(a)$ left-multiplied by $\Hm\Dm^{-1}$ and right-multiplied by $\Hm^\H$ \eqref{xxx.eq};
$(b)$ right-multiplied both sides of \eqref{yyy1.eq} by $(\Id_\nR+\Hm\Qm\Hm^\H)>\0$;
$(c)$ left-multiplied by $(\Hm^\H\Hm)^{-1/2}\Hm^\H$ and right-multiplied by $\Hm(\Hm^\H\Hm)^{-1/2}$ \eqref{yyy2.eq}, and finally applied the definitions:
\begin{align*}
\Km&\triangleq(\Hm^\H\Hm)^{1/2}\\
\Dmc&\triangleq\Dm^{-1}\\
\Fm&\triangleq\Km\Dmc\Km\\
\Rm&\triangleq\Km\Qm\Km
\end{align*}
The matrices $\Km,\Fm,\Rm$ are Hermitian, $\Km,\Fm$ are positive definite and $\Qm\ge\0$.
Since
$$\Fm\Rm=\Rm+\Rm^2=(\Fm\Rm)^\H=\Rm\Fm,$$
we have from \cite[Th.1.3.12]{horn} that $\Fm,\Rm$ are simultaneously unitarily diagonalizable, i.e.,
\begin{align*}
\Fm&=\Km\Dmc\Km=\Um\LamF\Um^\H\\
\Rm&=\Km\Qm\Km=\Um\LamR\Um^\H
\end{align*}
for some unitary matrix $\Um$ and diagonal matrices $\LamF,\LamR$.
Thus, we can rewrite \eqref{yyy3.eq} as
\begin{align}\label{Lam.eq}
  \LamF\LamR=\LamR+\LamR^2.
\end{align}
Given $\LamF$ we can see that $\LamR=(\LamF-\Id_\nT)_+$.
In fact, since $\LamF$ and $\LamR$ are diagonal matrices, we can consider any given diagonal position where the scalar equation is
$$\lambda_F\lambda_R=\lambda_R+\lambda_R^2=(\lambda_R+1)\lambda_R.$$
Since $\Qm\ge\0$ and $\Dmc>\0$, $\lambda_R\ge0$ and $\lambda_F\ge0$.
If $\lambda_F>1$, the equation has two possible solutions: $\lambda_R=0$ or $\lambda_R=\lambda_F-1$.
Since our goal is maximizing the mutual information, we choose the latter.
Otherwise, if $\lambda_F\le1$, $\lambda_R=0$ is the only possible solution.
Therefore, our chosen solution is $\lambda_R=(\lambda_F-1)_+$.
The extension to the matrix solution is direct.

Since $\Km$ is nonsingular, we have
\begin{align}
\Qm_\mathsf{opt}
&=\Km^{-1}\Um(\LamF-\Id_\nT)_+\Um^\H\Km^{-1}\non
&=\Km^{-1}(\Fm-\Id_\nT)_+\Km^{-1}\non
&=\Km^{-1}(\Km\Dmc\Km-\Id_\nT)_+\Km^{-1},
\end{align}
which proves eq.\ \eqref{Qopt.full.eq} of the Theorem and leads to the optimization problem \eqref{opt.full.problem}.
\end{IEEEproof}

\section{Proof of Theorem \ref{opt.singular.th}}\label{opt.singular.app}

\begin{IEEEproof}
A preliminary part of this proof is equivalent to that of Theorem \ref{opt.full.th} but the sequel is different because in this case the rank of $\Hm$ is smaller than $\nT$ and hence $\Hm^\H\Hm$ is singular.

As in the proof of Theorem \ref{opt.full.problem}, we obtain the equivalent optimization problem \eqref{opt.xxx.problem}.
Then, we have the following equivalent expression of \eqref{xxx.eq}:
\begin{align}\label{yyy4.eq}
\Fm\Rm=\Rm+\Rm^2
\end{align}
with a different definition of the matrices $\Fm,\Rm$:
\begin{align*}
\Dmc\triangleq\Dm^{-1},\qquad\Fm\triangleq\Hm\Dmc\Hm^\H,\qquad\Rm\triangleq\Hm\Qm\Hm^\H.
\end{align*}
Again, the matrices $\Fm,\Rm,\Qm$ are Hermitian positive semidefinite.
From \cite[Th.1.3.12]{horn}, $\Fm$ and $\Rm$ are simultaneously unitarily diagonalizable as
\begin{align*}
\Fm&=\Hm\Dmc\Hm^\H=\Um\LamF\Um^\H\\
\Rm&=\Hm\Qm\Hm^\H=\Um\LamR\Um^\H
\end{align*}
for some unitary matrix $\Um$ and diagonal matrices $\LamF,\LamR$.
Thus, we can rewrite \eqref{yyy4.eq} as
\begin{align}
  \LamF\LamR=\LamR+\LamR^2.
\end{align}
From the same arguments used after \eqref{Lam.eq}, we can see that $\LamR=(\LamF-\Id_\nu)_+$.
Then, from the reduced-size SVD $\Hm=\UmH\LamH\VmH^\H$, we obtain, for a given matrix $\Dmc$,
\begin{equation}\label{VQV.eq}
\LamH^{-1}\UmH^\H(\Hm\Dmc\Hm^\H-\Id_\nR)_+\UmH\LamH^{-1}=\VmH^\H\Qm\VmH.
\end{equation}

The proof of the Theorem follows.

As far as concerns the rank of the capacity-achieving covariance matrix $\Qm$, we recall two of the KKT equations \eqref{kkt.eq}:
\begin{subequations}
\begin{align}
\Hm^\H(\Id_\nR+\Hm\Qm\Hm^\H)^{-1}\Hm+\Mm&=\Dm\label{kkt1.eq}\\
\Mm\Qm&=\0\label{kkt2.eq}
\end{align}
\end{subequations}
From \eqref{kkt1.eq}, since $\Dm>\0$ (see Proof of Theorem \ref{opt.full.th}), so that $\rank(\Dm)=\nT$, and $\rank(\Hm^\H(\Id_\nR+\Hm\Qm\Hm^\H)^{-1}\Hm)=\nu$, we can apply inequality \cite[0.4.5.1]{horn} and obtain
\begin{align}\label{M.ineq}
\rank(\Mm)
&\ge\rank(\Dm)-\rank(\Hm^\H(\Id_\nR+\Hm\Qm\Hm^\H)^{-1}\Hm)\non
&=\nT-\nu.
\end{align}
Moreover, from \eqref{kkt2.eq} the matrices $\Mm,\Qm$ commute and hence~\cite[Th.1.3.12]{horn} we can write them as
$$\Mm=\Um\LamM\Um^\H,\Qm=\Um\LamQ\Um^\H.$$
Thus, we have $\LamM\LamQ=\0$ and then inequality \eqref{M.ineq} implies that
$$\rank(\Qm)=\rank(\LamQ)\le\nu.$$
Obviously, $\rank(\Qm)\ge1$ otherwise $\Qm$ would be the all-zero matrix.
This proves \eqref{Q.rank.eq} of Corollary \ref{Q.rank.th}.

Additionally, we note that the covariance matrix $\Qm$ is not determined only by $\Dmc$, which is specified by $\nu$ real parameters (the diagonal entries).
Since the capacity-achieving $\Qm$ has rank not greater than $\nu$ it can be specified by $(2\nT-\nu)\nu$ real parameters.
Conditions \eqref{VQV.eq} correspond to $\nu^2$ independent real equations so that the number of real variables in this optimization problem is given by $\nT+(2\nT-\nu)\nu-\nu^2=2(\nT-\nu)\nu+\nT$, which proves \eqref{complexity.eq} in Corollary \ref{complexity.th}.
\end{IEEEproof}

\section{Proof of Theorem \ref{unit.rank.singular.th}}\label{unit.rank.singular.app}

\begin{IEEEproof}
We can set $\nu=1,\UmH=\uv,\LamH=\|\Hm\|,\VmH=\vv$, with $\|\uv\|=\|\vv\|=1$, so that
$$\Hm=\|\Hm\|\uv\vv^\H.$$
Then, from Corollary \ref{Q.rank.th}, the capacity-achieving covariance matrix has unit rank and we set
$$\Qm=\qv\qv^\H.$$
Then, we notice that
\begin{align*}
\log\det(\Id_\nR+\Hm\Qm\Hm^\H)
&=\log\det(\Id_\nR+\|\Hm\|^2\uv\vv^\H\Qm\vv\uv^\H)\\
&=\log(1+\|\Hm\|^2|\vv^\H\qv|^2).
\end{align*}
Since
$$\vv^\H\qv=\sum_{i=1}^{\nT}|v_iq_i|\e^{\j\delta_i},$$
where $v_i\triangleq(\vv)_i,q_i\triangleq(\qv)_i,\delta_i\triangleq\angle q_i-\angle v_i$, we can see that
\begin{align*}
|\vv^\H\qv|^2
&=\sum_{i=1}^{\nT}\sum_{j=1}^{\nT}|v_iq_i||v_jq_j|\cos(\delta_i-\delta_j)\\
&\le\sum_{i=1}^{\nT}\sum_{j=1}^{\nT}|v_iq_i||v_jq_j|=\bigg\{\sum_{i=1}^{\nT}|v_iq_i|\bigg\}^2.
\end{align*}
The maximum is attained when $\delta_i$ is any constant so that we can set $\delta_i=0$.
This implies that the maximum value of $|\vv^\H\qv|$ is attained when $\angle q_i=\angle v_i$.

Once we found the phases of the $q_i$ we need the absolute values $w_i\triangleq|q_i|$, which can be found by considering the optimization problem
\begin{subequations}
\begin{align}
\min_{w_i,i=1,\dots,\nT}\quad & -\sum_{i=1}^{\nT}|v_i|w_i\\
s.t.                    \quad & -w_i\le0,i=1,\dots,\nT\\
                        \quad & w_i-\sqrt{P_i}\le0,i=1,\dots,\nT\\
                        \quad & \sum_{i=1}^{\nT}w_i^2-P_\mathsf{tot}\le0
\end{align}
\end{subequations}
The corresponding Lagrangian function is
\begin{align*}
\L(\wv)
&=-\sum_{i=1}^{\nT}|v_i|w_i-\sum_{i=1}^{\nT}\lambda_iw_i+\sum_{i=1}^{\nT}\mu_i(w_i-\sqrt{P_i})\\
&+\lambda\bigg\{\sum_{i=1}^{\nT}w_i^2-P_\mathsf{tot}\bigg\}
\end{align*}
The corresponding KKT equations are
\begin{subequations}
\begin{align}
-|v_i|-\lambda_i+\mu_i+2\lambda w_i=0\quad&i=1,\dots,\nT\label{kkt.rank1.eq1}\\
0\le w_i\le\sqrt{P_i}\quad&i=1,\dots,\nT\\
\lambda_iw_i=\mu_i(w_i-\sqrt{P_i})=0\quad&i=1,\dots,\nT\\
\sum_{i=1}^{\nT}w_i^2\le P_\mathsf{tot}\quad\label{kkt.rank1.eq4}\\
\lambda\bigg\{\sum_{i=1}^{\nT}w_i^2-P_\mathsf{tot}\bigg\}=0\quad\label{kkt.rank1.eq5}\\
\lambda\ge0,\lambda_i,\mu_i\ge0\quad&i=1,\dots,\nT\quad
\end{align}
\end{subequations}
\end{IEEEproof}
From \eqref{kkt.rank1.eq1} we get
$$w_i=\frac{|v_i|+\lambda_i-\mu_i}{2\lambda}$$
if $\lambda>0$.
The case $\lambda=0$ implies that the solution of \eqref{kkt.rank1.eq1} is undefined so that it must be discarded.
As a consequence, from \eqref{kkt.rank1.eq5}, the TP constraint \eqref{kkt.rank1.eq4} must be attained with equality.
Next, we can see from \eqref{kkt.rank1.eq1} that $w_i=0$ is not possible since $|v_i|>0,\lambda_i\ge0,\mu_i=0$.
If $w_i<\sqrt{P_i}$, then
$$w_i=\frac{|v_i|}{2\lambda}.$$
If $w_i=\sqrt{P_i}$, then
$$w_i=\frac{|v_i|-\mu_i}{2\lambda}.$$
These conditions can be summarized by setting
$$w_i=\min\bigg(\frac{|v_i|}{2\lambda},\sqrt{P_i}\bigg).$$
Defining the unknown $\alpha\triangleq1/(2\lambda)^2$, we have $w_i^2=\min(\alpha_i|v_i|^2,P_i)$ and get equation \eqref{alpha.eq}.
The lhs of this equation is a monotonically nondecreasing function of $\alpha$, which is equal to $0$ when $\alpha=0$ and is equal to $\sum_{i=1}^{\nT}P_i$ when $\alpha\to\infty$.
Thus, we have a single solution for $\alpha$, which can be found by Algorithm \ref{wf.algo}, which is based on rewriting \eqref{alpha.eq} in the equivalent form
$$\sum_{i=1}^{\nT}\min(\alpha,\rho_{\pi(i)})|v_{\pi(i)}|^2=P_\mathsf{tot}$$
where $\rho_i\triangleq P_i/|v_i|^2$ and $\pi$ is the permutation of the set $\{1,\dots,\nT\}$ such that $\rho_{\pi(i)}\le\rho_{\pi(i+1)}$ for $i=1,\dots,\nT-1$.
The solution is based on successive trials based on the assumption that $\alpha\le\rho_{\pi(i)}$ for $i=1,\dots,\nT$.

\begin{algorithm}
\caption{Algorithm for the solution of eq.~\eqref{alpha.eq}}\label{wf.algo}
\begin{algorithmic}[1]
\Procedure{Calculate-$\alpha$}{$v_i,P_i,P_\mathsf{tot},i=1,\dots,\nT$}
\State
Define $\rho_i\triangleq P_i/|v_i|^2$.
\State
Sort $\rho_i$ increasingly and let $\pi$ be the permutation such
\hspace*{4mm}
that $\rho_{\pi(i)}\le\rho_{\pi(i+1)},i=1,\dots,\nT-1$.
\For{$i=1,\dots,\nT$}
\State
$$\alpha=\frac{P_\mathsf{tot}-\sum_{k=1}^{i-1}P_{\pi(k)}}{\sum_{k=i}^{\nT}|v_{\pi(k)}|^2}$$
\If{$\alpha\le\rho_{\pi(i)}$}
\State\Return $\alpha$
\EndIf
\EndFor
\EndProcedure
\end{algorithmic}
\end{algorithm}


\end{document}